\documentclass[
aps,
prd,
reprint,
showpacs,
superscriptaddress,
longbibliography,
floatfix,
nobalancelastpage,
nofootinbib
]{revtex4-2}

\usepackage[bbgreekl]{mathbbol}
\usepackage{amsfonts}
\DeclareSymbolFontAlphabet{\mathbb}{AMSb}
\DeclareSymbolFontAlphabet{\mathbbl}{bbold}

\usepackage{linenoAMS}
\usepackage{amsmath}
\usepackage{siunitx}
\usepackage[colorlinks=true, citecolor=RoyalBlue,
linkcolor=BrickRed, urlcolor=ForestGreen]{hyperref}
\usepackage[all]{hypcap}

\usepackage{enumitem}
\usepackage{autofigs9}
\usepackage{braket}
\usepackage{comment}
\usepackage{mathtools}

\graphicspath{{figures/}}
\usepackage[usenames, dvipsnames, svgnames, table]{xcolor}

\usepackage{graphicx}
\usepackage{xfrac}
\usepackage{ulem}
\usepackage{cancel}

\usepackage[capitalise]{cleveref}

\begin{document}
\title{Experimental demonstrations of alignment and mode matching in optical cavities with higher-order Hermite-Gauss modes}

\author{Liu Tao}
\email{liu.tao@ligo.org}
\affiliation{University of Florida, 2001 Museum Road, Gainesville, Florida 32611, USA}
\author{Paul Fulda}
\affiliation{University of Florida, 2001 Museum Road, Gainesville, Florida 32611, USA}

\date{\today}

\begin{abstract}
Higher-order spatial laser modes have recently been investigated as candidates for reducing test-mass thermal noise in ground-based gravitational-wave detectors such as advanced LIGO. In particular, higher-order Hermite-Gauss (HG) modes have gained attention within the community for their more robust behaviors against random test-mass surface deformations and stronger sensing and control capacities. In this letter we offer experimental investigations on various aspects of HG mode interferometry. We have generated purified HG modes up to the 12-th order $\mathrm{HG}_{6,6}$ mode, with a power conversion efficiency of 38.8\% and 27.7\% for the $\mathrm{HG}_{3,3}$ and $\mathrm{HG}_{6,6}$ modes respectively. We demonstrated for the first time the misalignment and mode mismatch-induced power coupling loss measurements for HG modes up to the $\mathrm{HG}_{6,6}$. We report an excellent agreement with the extended numerical power loss factors that in the ``small power loss'' region converge to $2n+1$ or $n^2+n+1$ for a misaligned or mode mismatched $\mathrm{HG}_{n,n}$ mode. We also demonstrated the wavefront sensing (WFS) signal measurement for HG modes up to the $\mathrm{HG}_{6,6}$. The measurement result is accurately in accordance with theoretical WFS gain $\beta_{n,n-1}\sqrt{n} + \beta_{n,n+1}\sqrt{n+1}$ for an $\mathrm{HG}_{n,n}$ mode, with $\beta_{n,n-1}$ being the beat coefficient of the adjacent $\mathrm{HG}_{n,n}$ and $\mathrm{HG}_{n-1,n}$ modes on a split photodetector. 
\end{abstract}

\maketitle

Laser beams with higher-order spatial transverse modes (HOMs) are actively investigated as a beneficial alternative for the currently used fundamental Gaussian laser beam to reduce the thermal noise of interferometric gravitational-wave (GW) detectors~\cite{Mours_2006, Vinet_2007, PhysRevD.82.042003}. This offers valuable improvements on the detector sensitivities at frequency bands that are limited by the thermal noise, such as at frequencies around 100 Hz for the current GW detectors such as advanced LIGO (aLIGO) and advanced Virgo~\cite{aLIGO, PhysRevD.102.062003, AdVirgo, Acernese_2023}, or at around 10 Hz for the next-generation detectors such as Cosmic Explorer and Einstein Telescope~\cite{CE, ET}. Unlike the previously favored higher-order Laguerre-Gauss modes with their susceptibilities against realistic mirror surface imperfections~\cite{PhysRevD.84.102002, Sorazu, PhysRevD.82.012002}, higher-order Hermite-Gauss (HG) modes with their symmetry between the tangential and sagittal components, can be made compatible with the current mirror polishing techniques with the deliberate addition of astigmatism~\cite{PhysRevD.102.122002, PhysRevD.103.042008}. It has also been demonstrated by the authors that higher-order HG modes offer stronger alignment and mode-matching sensing capacities in different sensing schemes~\cite{PhysRevD.108.062001}, which helps maintain optimal alignment and mode-matching working states in the interferometer and limits the power coupling loss scattered to other non-resonant modes~\cite{Jones, Tao:21}. Higher-order HG modes have also been experimentally investigated with their high-efficiency generation at higher power~\cite{10.1063/5.0137085}, and the compatibility with the squeezed light generation techniques for quantum noise reduction~\cite{PhysRevLett.128.083606, PhysRevLett.129.031101}. 

Beyond the thermal noise benefit in GW detectors, higher-order modes are also a prominently active area of study in a variety of applications, such as a finer characterization of topologically complex electronic matter~\cite{LG_Lee}, high-resolution imaging for object identification~\cite{PhysRevLett.110.043601}, improved-precision small-displacement measurements~\cite{10.1063/1.4869819}, and high-efficiency multimode quantum communication and information processing~\cite{PhysRevLett.98.083602}.

This paper focuses on the experimental aspects of higher-order HG mode interferometry, verifying the misalignment and mode mismatch-induced power coupling loss scaling relations, and the compatibility with the wavefront sensing (WFS) and control techniques that are currently implemented in aLIGO. We start with higher-order HG mode generation that uses a computer-controlled liquid-crystal-on-silicon spatial light modulator (SLM). A Gaussian beam incident on the SLM screen picks up a ``checkerboard-shaped'' phase profile that resembles the phase front of higher-order $\mathrm{HG}_{n,n}$ modes at the waist. The converted beam is purified with a pre-mode-cleaner (PMC) cavity and is used to pump a downstream cavity. With the mode matching to the PMC optimized for the $\mathrm{HG}_{3,3}$ mode, we report a power conversion efficiency of 38.8\% for the $\mathrm{HG}_{3,3}$ mode that is limited by the diffraction and absorption loss of the SLM. 

Previously we have shown through an analytical calculation that higher-order HG modes suffer more power coupling loss when,
for example, coupling into the eigenmode of an optical cavity, due to the mode scattering induced by misalignment and mode mismatch. Specifically, in the ``near-perfect'' alignment and mode matching states, an $\mathrm{HG}_{n,n}$ mode suffers $2n+1$ times more power loss than the fundamental $\mathrm{HG}_{0,0}$ mode for the same alignment state, and it suffers $n^2+n+1$ times more power loss with the same mode matching condition~\cite{Tao:21}. In this letter, we pushed beyond the ``near-perfect" beam perturbation region, and investigated the power loss behavior for HOMs with increasing misalignment and mode mismatch. We found that with a finite amount of misalignment and mode mismatch, the power loss scaling relations start to deviate from the analytical results that only take the nearest sets of scattered modes into the power loss consideration. More accurate and robust numerical results that use no modal approximation for the misalignment and mode mismatch-induced power loss for HOMs up to the $\mathrm{HG}_{6,6}$ are shown beside the experimental results, and we report excellent agreements. 

We have also previously shown analytically and with simulations using \textsc{FINESSE}~\cite{finesse, pykat, brown2020pykat} that despite the fact that higher-order HG modes suffer more power loss from a given amount of misalignment and mode mismatch, they also provide stronger alignment and mode mismatch sensing signals~\cite{PhysRevD.108.062001}. This provides better sensing and control capacities for higher-order HG modes over misalignment and mode mismatch, which could be used to mitigate their excessive power coupling losses. In particular, in the wavefront alignment sensing scheme, an $\mathrm{HG}_{n,n}$ mode generates a sensing signal that is $\beta_{n,n-1}\sqrt{n}+\beta_{n,n+1}\sqrt{n+1}$ times stronger than for the $\mathrm{HG}_{0,0}$ mode, where $\beta_{n,n-1}$ for instance is the beat coefficient between the adjacent $\mathrm{HG}_{n,n}$ and $\mathrm{HG}_{n-1,n}$ modes on a split photodetector~\cite{PhysRevD.108.062001}. In this letter, we report for the first time an experimental demonstration of the wavefront alignment sensing signals for HOMs up to the $\mathrm{HG}_{6,6}$ mode. We report an excellent agreement of the WFS gain measurement with the theoretical prediction.

\begin{figure}[t]
    \centering
    \includegraphics[width=\linewidth]{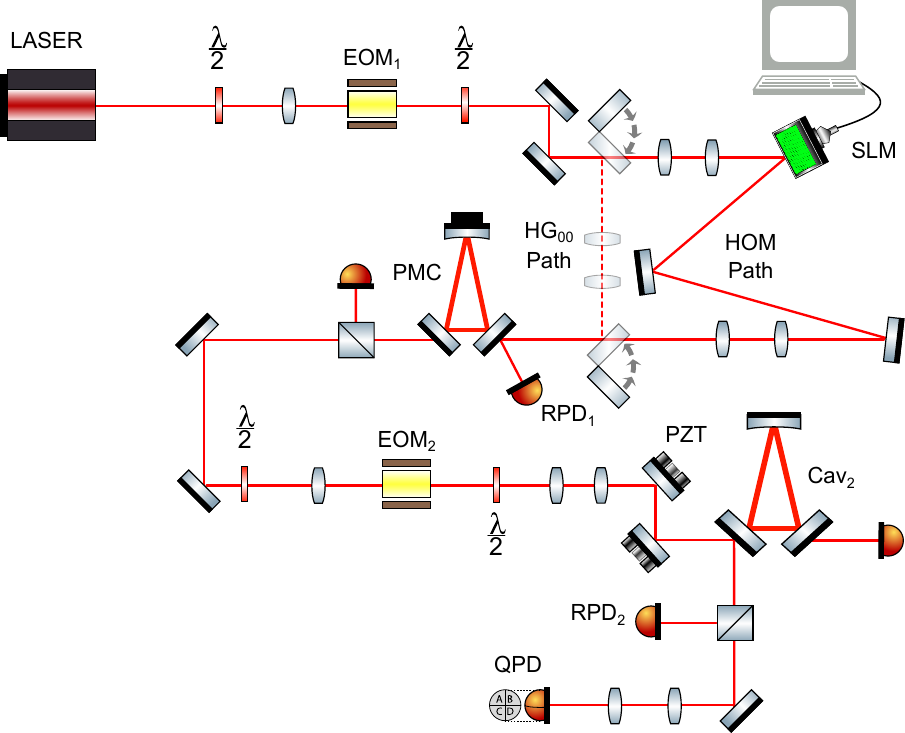}
    \caption{Experiment layout}
    \label{fig: labsetup}
\end{figure}

\textbf{Basic Setup} 
In order to generate higher-order HG modes, we implemented the 1920x1152 XY Phase Series of liquid crystal on silicon spatial light modulator (SLM) from Meadowlark Optics. The ``Checkerboard-shaped'' phase maps were applied to the SLM, with the accumulated phase being either $0^{\circ}$ or $180^{\circ}$ to mimic the phase front of a higher-order HG mode at its waist. The converted beam reflected off the SLM was mode-matched to a pre-mode cleaner cavity, labeled as ``PMC" in Fig.~\ref{fig: labsetup}. The laser was locked to a particular HOM eigenmode of the PMC through the Pound-Drever-Hall (PDH) technique by actuating the laser piezo and the crystal temperature. The purified HOM beam transmitted through the PMC was used to pump a subsequent cavity, labeled as ``$\mathrm{Cav_{2}}$" in Fig.~\ref{fig: labsetup}. The laser was then locked to the same HOM eigenmode of ``$\mathrm{Cav_{2}}$" by sending the PDH control signal from the second control loop to the end mirror piezo of the PMC, which ``drags'' the laser frequency along through the first closed control loop to match the resonance frequency of the second cavity. $\mathrm{Cav_{2}}$ acts as a mode reference cavity that was used to demonstrate the misalignment and mode mismatch power coupling losses, as well as the wavefront alignment sensing for HOMs. The wavefront sensing measurement was made by the quadrant photodetector (QPD) in the reflection of $\mathrm{Cav_{2}}$. It detects and demodulates the beat signal between the radio-frequency (RF) sideband field generated from the second electro-optic phase modulator, labeled as ``$\mathrm{EOM_{2}}$'', and the misalignment-induced carrier
frequency offset modes.

We also have a separate beam path that bypasses the SLM. This ``$\mathrm{HG}_{0,0}$-path", indicated as the dashed lines in Fig.~\ref{fig: labsetup}, functions as a reference for the impedance matching condition of the PMC, and was used to characterize the conversion efficiency of the HOM generation.

\textbf{HOM Conversion Efficiency}
We want to optimize the HOM conversion process, generating as much purified HOMs as possible from a given fundamental mode input beam. The conversion efficiency is quantified by comparing the transmitted power through the PMC in a particular HOM to the maximum power coupling to the PMC for the fundamental mode, measured by the $\mathrm{HG}_{0,0}$-path. With optimal alignment, the power mode mismatch for the $\mathrm{HG}_{0,0}$-path is 1.8\%, by comparing the second-order modes power to the fundamental mode. The power in the $\mathrm{HG}_{0,0}$ mode transmitted through the PMC is 255 mW. This indicates that with perfect mode matching, the maximum coupling power for the PMC would be 255/(1-0.018) = 260 mW, which is used as a reference to estimate the HOM conversion efficiency.
In this $\mathrm{HG}_{0,0}$-path, the incident power on the PMC (the same as the power incident on the SLM for the ``HOM-path'') is 299 mW, indicating that the PMC is not 100\% transmissive. 

The mode matching for the HOM path was optimized for the $\mathrm{HG}_{3,3}$ mode. The optimal phase map size for generating an $\mathrm{HG}_{3,3}$ mode compared to the beam size on the SLM is around 0.4 according to our numerical calculation. Optimizing the mode matching between the converted beam and the PMC we were able to get 101 mW purified $\mathrm{HG}_{3,3}$ mode, indicating the conversion efficiency for generating the $\mathrm{HG}_{3,3}$ mode is 101/260 = 38.8\%. The coupling power and conversion efficiency for HOMs up to $\mathrm{HG}_{6,6}$ are listed as $\mathrm{P_{trans}}$ and $\eta_{1}$ in Table.~\ref{tab: HOMefficiency}. The mode matching was not re-optimized for the other HOMs, which indicates a higher mode mismatch power loss for them.

\begin{table}[htbp]
\centering
\caption{The power transmitted through the PMC $\mathrm{P_{trans}}$ and the conversion efficiency $\eta_{1}$ and $\eta_{2}$ for HOMs up to $\mathrm{HG}_{6,6}$. }
\begin{tabular}{c|c|c|c|c|c|c}
\hline \hline 
HOM & $\mathrm{HG}_{1,1}$&$\mathrm{HG}_{2,2}$&$\mathrm{HG}_{3,3}$&$\mathrm{HG}_{4,4}$&$\mathrm{HG}_{5,5}$&$\mathrm{HG}_{6,6}$\\
\hline
$\mathrm{P_{trans}}$ (mW) & 104 & 97 & 101 & 87 & 75 & 72\\
\hline 
$\eta_{1}$ (\%) & 40.0 & 37.3 & 38.8 & 33.5 & 28.8 & 27.7\\
\hline 
$\eta_{2}$ (\%) & 54.6 & 51.0 & 53.0 & 45.7 & 39.4 & 37.8\\
\hline
\hline
\end{tabular}
\label{tab: HOMefficiency}
\end{table}

There is also a substantial amount of power lost on the SLM itself from the absorption and diffraction, due to the fill factor. Eight diffraction orders are visible in each direction, with the total diffraction power loss being 21 mW. With 299 mW of light incident on the SLM, there is only 219 mW of power in the main diffraction order that incident on the PMC. This suggests that if we do not factor the SLM loss into our consideration of the HOM conversion efficiency, and only consider our method of using the ``checkerboard-shaped'' phase profiles and purifying with a PMC, the conversion efficiency for the $\mathrm{HG}_{3,3}$ mode would be $101/260\cdot (299/219)=53.0$ \%. This is rather close to the theoretical maximum conversion efficiency of 56\% for the $\mathrm{HG}_{3,3}$ mode according to our numerical calculation. The corresponding efficiency for HOMs up to $\mathrm{HG}_{6,6}$ is shown as $\eta_{2}$ in Table.~\ref{tab: HOMefficiency}. 

\begin{figure}[b]
    \centering
    \includegraphics[width=\linewidth]{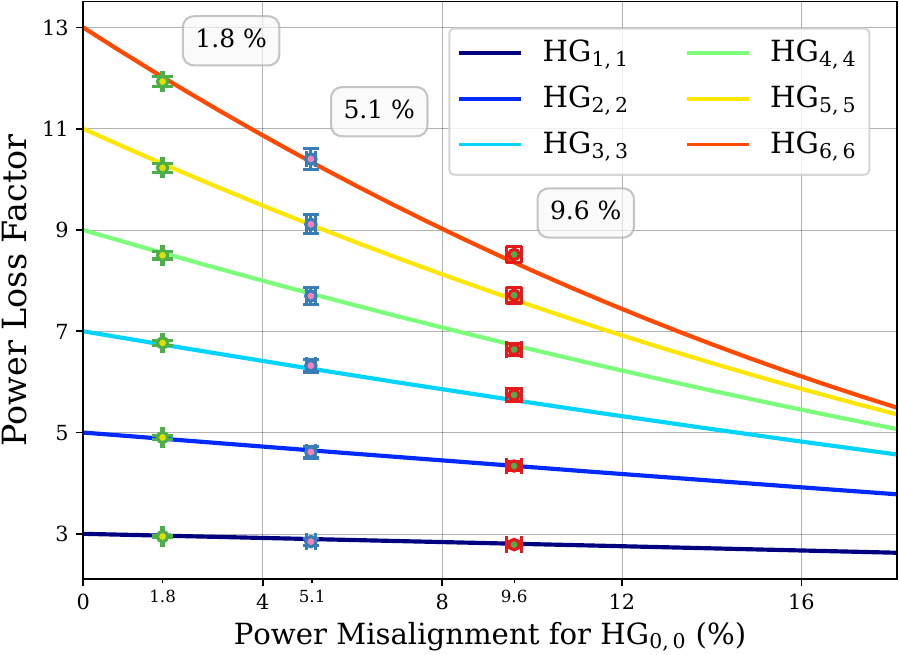}
    \caption{Misalignment induced power loss factors for HOMs with an increasing amount of misalignment. }
    \label{fig: misalignmentloss}
\end{figure}

\begin{figure}[b]
    \centering
    \includegraphics[width=\linewidth]{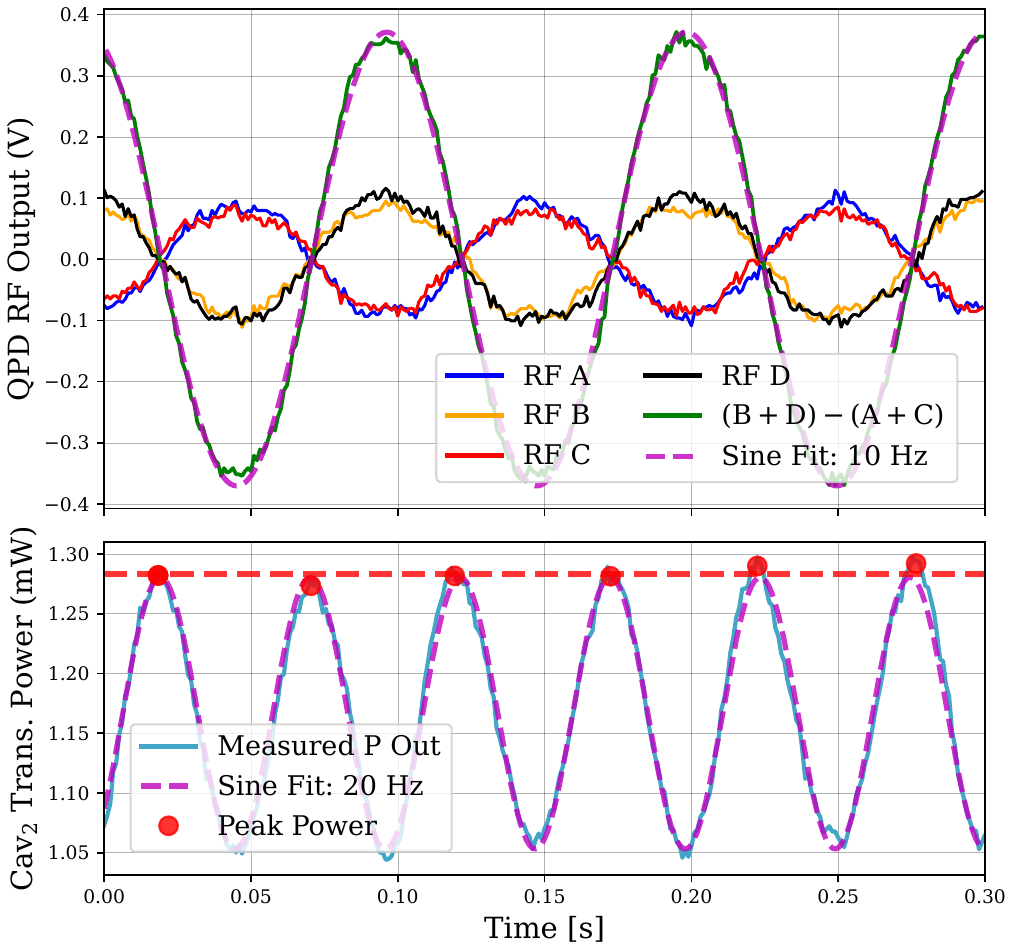}
    \caption{WFS measurement for the $\mathrm{HG}_{3,3}$ mode. Top: RF demodulated output from each of the four QPD quadrants, and the WFS signal by subtracting the right quadrants to the left; Bottom: the $2f$ transmitted PD signal.}
    \label{fig: WFSMeasurment_HG33}
\end{figure}

\textbf{Misalignment Induced Power Coupling Loss} With both cavities locked to the same HOM, the piezo mirror before $\mathrm{Cav_{2}}$ (labeled as ``PZT") was actuated with a sinusoid at $f=10$ Hz. When the amplitude of the drive signal is small, and it is centered around the ideal alignment, the transmitted photodiode (PD) signal should also be a sinusoid, at twice the drive frequency (20 Hz), an example of which is shown at the bottom panel of Fig.~\ref{fig: WFSMeasurment_HG33}. This is because when the misalignment is small, only the scattered modes that are separated from the original mode by one mode order contribute to the power loss, which scales as $\sin^2(2\pi \cdot f t) = \left(1-\cos(2\pi \cdot 2f t)\right)/2$. As the misalignment gets larger, the scattered modes that are separated from the original mode by $\textit{more than}$ one mode order must be considered for the total power loss. This produces a $2f$ PD signal that is no longer a pure sinusoid, due to the high-frequency contributions from the large offset modes.

We can then infer the misalignment-induced power loss from the $2f$ signal, by normalizing the peak-to-peak amplitude of the signal with respect to its peak value. The two cavities were locked to the $\mathrm{HG}_{0,0}$, $\mathrm{HG}_{1,1}$, ..., $\mathrm{HG}_{6,6}$ modes. The power loss for each HOM normalized by the power loss for the $\mathrm{HG}_{0,0}$ mode (the power loss factor for HOMs) was measured with three different alignment states, corresponding to 1.8\%, 5.1\%, and 9.6\% power loss for the $\mathrm{HG}_{0,0}$ mode. They are shown in the green dots, the blue dots, and the red dots respectively in Fig.~\ref{fig: misalignmentloss}. The x and y error bars from the figures were obtained from a repeat measurement of five times. 

We have also numerically calculated the misalignment-induced power loss factor for HOMs beyond the ``near-perfect" alignment region. This was achieved by calculating the overlap of the 2D beam amplitude arrays of the unperturbed beam and the offset beam. The numerical results are shown in the lines in Fig.~\ref{fig: misalignmentloss} for different HOMs. We see that near perfect alignment, the power loss factors for HOMs scale as $2n+1$, with $n$ being the mode index~\cite{Tao:21}. However, the power loss factors decrease as the amount of misalignment increases, due to the convergence of the mode overlap coefficients at large beam perturbations, meaning that the contributions from the scattered modes that are offset from the original mode by more than one mode order become more and more significant. The numerical results agree excellently with the measurement, for all three different alignment states.

\begin{figure}[h!tbp]
    \centering
    \includegraphics[width=\linewidth]{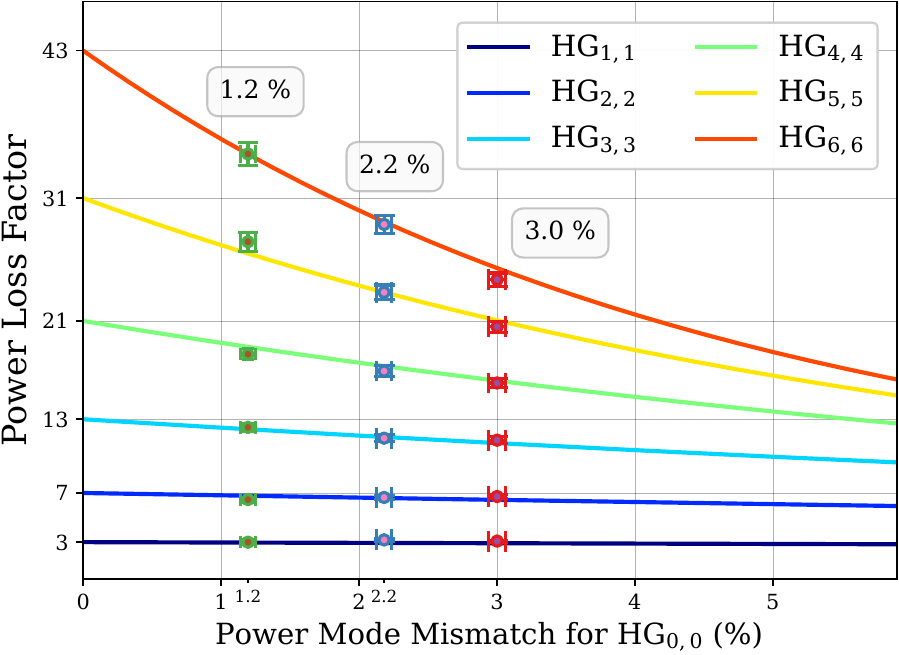}
    \caption{Mode mismatch induced power loss factors for HOMs with an increasing amount of mode mismatch.}
    \label{fig: modemismatchloss}
\end{figure}

\textbf{Mode Mismatch Induced Power Coupling Loss} The mode mismatch-induced power loss factor for HOMs was measured by comparing the power in the mode resonances scattered from the original mode by even mode orders to the summation of the scattered power and the power in the original mode. Three different mode matching states were obtained, at 1.2\%, 2.2\%, and 3.0\% power mode mismatch for the $\mathrm{HG}_{0,0}$ mode. The corresponding measured power loss factors are shown in green dots, blue dots, and red dots respectively in Fig.~\ref{fig: modemismatchloss}.

\begin{figure*}[t]
    \centering
    \includegraphics[width=\linewidth]{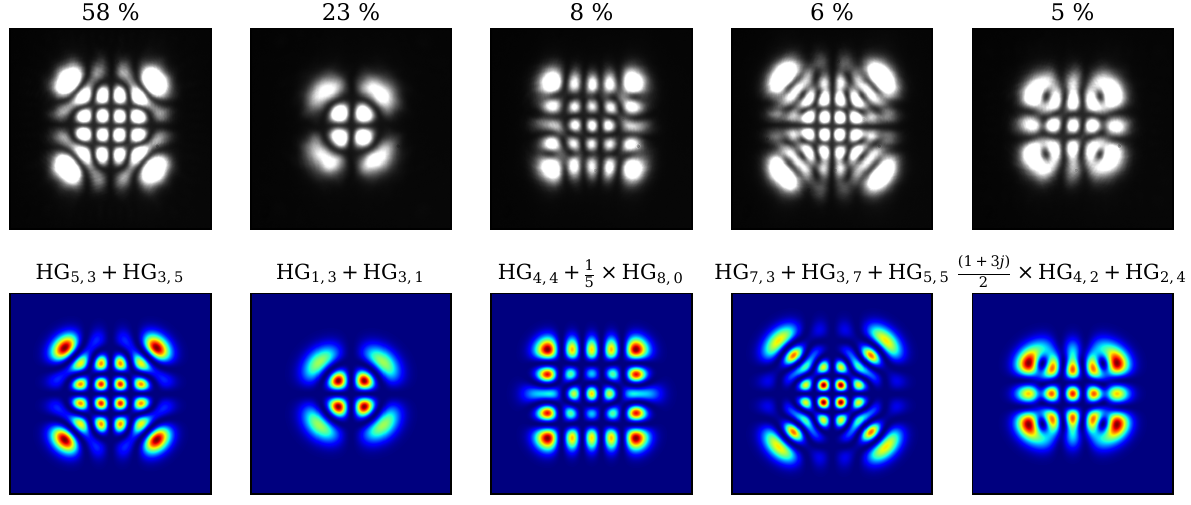}
    \caption{Top: Measured intensity images for the five mode-mismatch scattered modes for the $\mathrm{HG}_{3,3}$ mode (3\% power loss for $\mathrm{HG}_{0,0}$), with their relative power in each scattered mode; Bottom: Corresponding modal estimation for the scattered modes.}
    \label{fig: HG33ModeMismatchScattering}
\end{figure*}

The mode mismatch-induced power loss factors for HOMs, as we move away from the ``near-perfect'' mode-matching state, have also been numerically calculated, similar to the misalignment case. The numerical results are shown in the lines in Fig.~\ref{fig: modemismatchloss} for different HOMs. We see that when we are near the perfect mode matching state, the power loss factors for HOMs scale as $n^2+n+1$, with $n$ being the mode index~\cite{Tao:21}. And the power loss factors with finite mode mismatch decrease as the mode matching gets worse. The numerical results are also in good agreement with the measurement.

We also noticed a slight discrepancy between the measured mode mismatch-induced power loss factor and the corresponding numerical result in Fig.~\ref{fig: modemismatchloss},  particularly for higher-order $\mathrm{HG}_{5,5}$ and $\mathrm{HG}_{6,6}$ modes at 3\% mode mismatch power loss for the $\mathrm{HG}_{0,0}$ mode. This is most likely due to an undercounting of the scatter modes, since the number of scattered modes gets large at large mode mismatch, especially for higher-order modes. For instance, for the $\mathrm{HG}_{3,3}$ mode at 3\% mode mismatch, there were five significant scattered modes identified for the total power loss, as shown on the top row in Fig.~\ref{fig: HG33ModeMismatchScattering}, ranked by their contribution to the total power loss. We were also able to identify the modal contents for each scatter mode resonance, shown on the bottom row in Fig.~\ref{fig: HG33ModeMismatchScattering}. For instance, we see that for the $\mathrm{HG}_{3,3}$ mode at 3\% mode mismatch, the scattered mode resonance that contributes the most to the total power loss is the odd 8th order modes and the odd 4th order modes, which are 2 mode orders away from the original 6th order $\mathrm{HG}_{3,3}$ mode. This is followed by the even 8th order modes, 10th order modes, which are separated by 4 mode order from the original $\mathrm{HG}_{3,3}$ mode, and then the even 6th order modes.

\textbf{Wave-front Alignment Sensing Gain} For the WFS signal measurement for HOMs, the PZT mirror before $\mathrm{Cav}_{2}$ was actuated with a sinusoid at $f=10$ Hz, and the QPD in the reflection was placed at the image plane of the PZT mirror to reduce any residual radio-frequency amplitude modulation (RFAM) effect caused by the beam spot movement on the QPD. 

\begin{figure}[htbp]
    \centering
    \includegraphics[width=\linewidth]{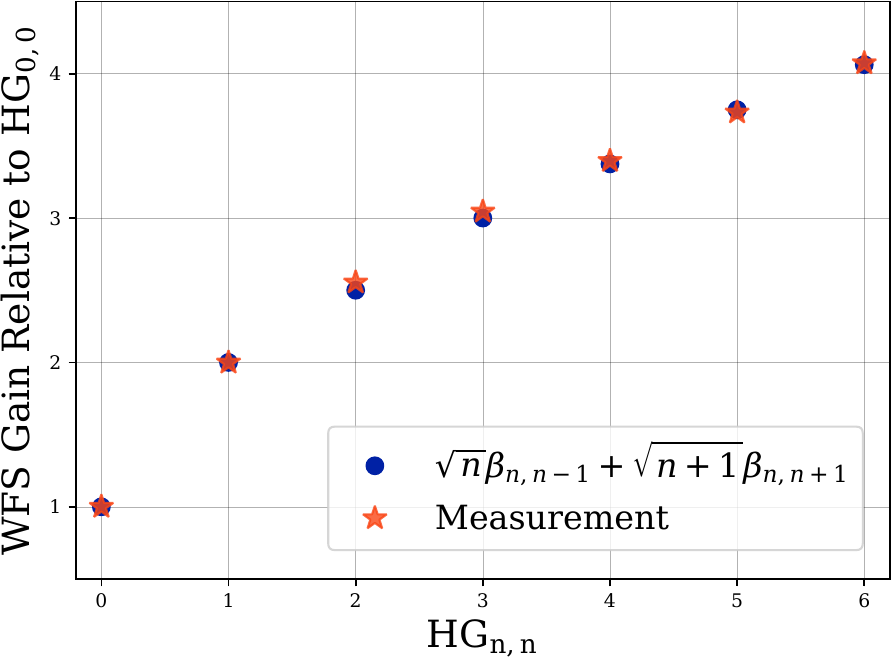}
    \caption{The WFS gain measurements for HOMs up to $\mathrm{HG}_{6,6}$.}
    \label{fig: WFSGainHOM}
\end{figure}

The WFS signal experiment was carried out with both cavities locked to an HG mode from the $\mathrm{HG}_{0,0}$ to $\mathrm{HG}_{6,6}$ mode. For instance, the WFS measurement for the $\mathrm{HG}_{3,3}$ mode is shown in Fig.~\ref{fig: WFSMeasurment_HG33}. We obtain a Sine-like RF demodulated signal from each of the four quadrants of the QPD. They have the same amplitude, with a $180^{\circ}$ phase difference between the left and right quadrants. The WFS signal from a split photodetector was obtained by subtracting the demodulated signals from the right quadrants of the QPD to the left, which is also a sinusoid with frequency $f$. The $2f$ transmitted PD signal is shown at the bottom of Fig.~\ref{fig: WFSMeasurment_HG33}. To get the WFS signal amplitude for each HOM, we normalize the amplitude of the RF demodulated signal from the split photodetector by the peak intensity of the transmitted power. The WFS signal amplitude for each HOM, normalized by the amplitude for the $\mathrm{HG}_{0,0}$ mode (WFS gain), is shown in Fig.~\ref{fig: WFSGainHOM}. We have an excellent agreement between the measured WFS gain for HOMs up to $\mathrm{HG}_{6,6}$ with the theoretical result $\sqrt{n}\beta_{n,n-1} + \sqrt{n+1}\beta_{n,n+1}$, with $\beta_{n,n-1}$ for instance being the beat coefficient between the $\mathrm{HG}_{n,n}$ and $\mathrm{HG}_{n-1,n}$ on the split photodetector.

\textbf{Conclusion}
With the ``checkerboard-shaped'' phase profiles applied to the SLM and mode purification with the PMC, we were able to generate higher-order HG modes up to the $\mathrm{HG}_{6,6}$ with high purity and efficiencies. For instance, the conversion efficiency for the $\mathrm{HG}_{3,3}$ and $\mathrm{HG}_{6,6}$ mode is 38.8\% and 27.7\% respectively, with the mode matching designed for the optimal conversion for the $\mathrm{HG}_{3,3}$ mode. We then used the purified HOM to pump a subsequent cavity to demonstrate their power coupling loss factors induced by misalignment and mode mismatch, as well as the wavefront alignment sensing signals. We extended the numerical power coupling loss factors for HOMs beyond the ``near-perfect'' alignment and mode-matching region~\cite{Tao:21}. And we were able to demonstrate excellent agreement between the experimentally measured power loss factors and the extended numerical results in three different misalignment and mode mismatch states. We also experimentally demonstrated the WFS signals for higher-order HG modes for the first time, which were shown accurately in accordance with the corresponding analytical WFS gains~\cite{PhysRevD.108.062001}.

\section*{Acknowledgments}
This work was supported by National Science Foundation grants PHY-1806461 and PHY-2012021. The authors would also like to thank Harold Hollis for the design and build of the quadrant photodiode used in the measurement.

\nocite{apsrev42Control}
\bibliographystyle{apsrev4-2-trunc}
\bibliography{paper}

\end{document}